# Top Down Approach to find Maximal Frequent Item Sets using Subset Creation

Jnanamurthy HK, Vishesh HV, Vishruth Jain, Preetham Kumar, Radhika M. Pai
Department of Information and Communication Technology
Manipal Institute of Technology, Manipal University, Manipal-576104, India
jnanamurthy.hk@gmail.com

***Abstract***--*Association rule has been an area of active research in the field of knowledge discovery. Data mining researchers had improved upon the quality of association rule mining for business development by incorporating influential factors like value (utility), quantity of items sold (weight) and more for the mining of association patterns. In this paper, we propose an efficient approach to find maximal frequent itemset first. Most of the algorithms in literature used to find minimal frequent item first, then with the help of minimal frequent itemsets derive the maximal frequent itemsets. These methods consume more time to find maximal frequent itemsets. To overcome this problem, we propose a navel approach to find maximal frequent itemset directly using the concepts of subsets. The proposed method is found to be efficient in finding maximal frequent itemsets.*

## 1. INTRODUCTION

With the popularization of computer and development of Database Technology, more and more data are stored in large databases. Obviously, it is impossible to find useful information without using efficient methods. Data Mining (DM)[1] techniques have emerged as a reflection of this request. Association rules mining, an important research direction aims to find out the dependence among multiple domains based on a given degree of support and credibility.

Association rules mining process is divided into two steps. The first step is to find the frequent item-sets whose support degree is larger than the initial support degree from the transaction database; the second step is to generate the rules of value from the frequent item-sets, and the acquisition of frequent item-sets is the key step during mining association rules procedure. In 1993, R. Agrawal first promoted an association rule mining algorithm named Apriori Algorithm[2].This algorithm's basic idea is to identify all the frequent sets whose support is greater than minimum support. Frequent item generates strong association rule, which must satisfy minimum support and minimum confidence. An Apriori idea is a brief description of the core algorithm is that has two key steps: the connecting step and the pruning step [3].

- Connecting step: In order to identify the L(k) (a frequent k set), a candidate k-items (C(k)) can be generated by L(k)-1 and its connections, which elements of L(k)-1 can be connected.

- Pruning step: C(k) is a superset of L(k) whose members may or may not be frequent, but all the frequent sets are included in C(k). If scanning database, each count of a candidate in C(k) can be determined, also L(k) (frequent candidates whose count is not less than the minimum support count). However, C(k) may be large, its calculation amount also be lots. For compression of C(k) the Apriori may be used: any non-frequent (k-1) items can not be subsets of frequent k-items. Therefore, if (k-1) items of a candidate k-items is not in L(k), then the candidate cannot be frequent, which can be deleted from C(k).

Subsequent researchers have given a lot of improvement for the AA. However, all of these improved algorithms have the following problems in varying degrees. The first problem is that algorithms need more time complexity to produce the candidate frequent item-sets. And the second is that algorithms have to scan the transaction database many times to do the pattern-matching for candidate frequent item-sets. These two issues are both the hotspots and difficulties during current research on mining association rules. In our paper, we promote a faster and more efficient algorithm based on the classical AA.

## 2. BASIC CONCEPTS.

Data Mining is a method that extracts some kind of information knowledge which cannot be discovered easily, but contains certain regularity from the massive primary data [4]. Let *I* be a set of items and *D* a database of transactions. Every transaction is a set of distinct items (itemset) from *I*. An itemset with *k* items is referred to as a *k-itemset*. The *support of an itemset X*, denoted as $\sigma(X)$, is the total number of transactions in which that itemset occurs as a subset. A second formal definition for the support of an itemset *X* is given by Agrawal. An itemset *X* has a support of *s* if *s%* of transactions in *D* contains *X* as a subset. This second formal definition is somewhat more rigorous, as it emphasizes that the maximum support of an ite set cannot exceed the total number of transactions in *D*. An itemset is called *frequent* if its support is greater than a user-defined *minimum support* value. A frequent *k-itemset X* is *maximal* if no other *k'-itemset* (where $k < k'$) contains *X* as a subset.

An *association rule* is an expression $X \Rightarrow Y$, where *X* and *Y* are disjoint itemsets. An important note is that an association rule must not be considered not only as an implication, but rather as a coexistence of the two itemsets. The *support* of an association rule is given by the support of the $X \cup Y$ itemset. The *confidence* of an association rule is the conditional Probability that a transaction contains *Y*, given that it contains *X*. The confidence is computed using the formula $c(X \Rightarrow Y) = \sigma(X \cup Y) / \sigma(X)$. *Minimum confidence* of a rule is a user defined value. An association rule is *strong* if it has a support greater than minimum support value and confidence greater than the minimum confidence value.

## 3. THEORETICAL BACKGROUND

**Association rules:** Association rules are statements of the form {X1,X2….Xn}→Y , meaning that if we find all of X1,X2….Xn in the market basket, then we have a good chance of finding Y. We normally would search only for rules that had confidence above a certain threshold. We may also ask that the confidence be significantly higher than it would be if items were placed at random into baskets.

**Frequent itemsets:** In many (but not all) situations, we only care about association rules or causalities involving sets of items that appear frequently in baskets. For example, we cannot run a good marketing strategy involving items that no one buys anyway. Thus, much data mining starts with the assumption that we only care about sets of items with high support; i.e., they appear together in many baskets. We then find association rules or causalities only involving a high-support set of items i.e., {X1. . .Xn ,Y } must appear in at least a certain percent of the baskets, called the support threshold.

What is the use of learning association rules?

- With the development of e-commerce and logistics, online shopping plays an increasingly important role in people's life. Some well-known e-commerce site gets lots of benefits from mining association rules. These online shopping sites use mining association rules to get useful information from the huge database, and then set the commodity in a bundle that the customer intends to purchase together. And there are also some shopping sites which use them to set the appropriate cross-selling, where the customer who bought one product will see other related commodities advertised. [5]

- Also we are familiar with Amazon; they use association mining to recommend you the items based on the current item you are browsing/buying.
- Another application is the Search engines where after you type in a word, it searches for frequently associated words that the user types after that particular word. [6]

# 4. PROPOSED METHOD

Fig.1 shows activity diagram of MFIF method to find maximal frequent item first. Instead of finding minimal frequent itemset first, we developed a new efficient method to find maximal frequent itemset first.

**Procedure:**
**Step1:** Count the number of items present in each transaction and put in an array a[ ].
**Step2:** Find the transactions having maximum items (max) in the array a[ ].
**Step3:** If Count (max (a[ ]) ) ≥ min_sup then transfer those transactions to an another array arr[ ][ ],else find subsets.
**Step4:** Compare each transaction in arr[ ][ ] with other transactions.
**Step5:** Take a Counter C and increase the counter if we found similar itemsets in arr[ ][ ].
**Step6:** If {C ≥ min_sup} then itemset will be the most frequent itemset.
**Step7:** if C<min_sup then find the subsets of all transactions and store it in an array sub[ ][ ].
**Step8:** max = max-1.
**Step9:** add the transactions of sub[ ][ ] to arr[ ][ ].
**Step10:** Repeat from **step3** until frequent itemset is found.

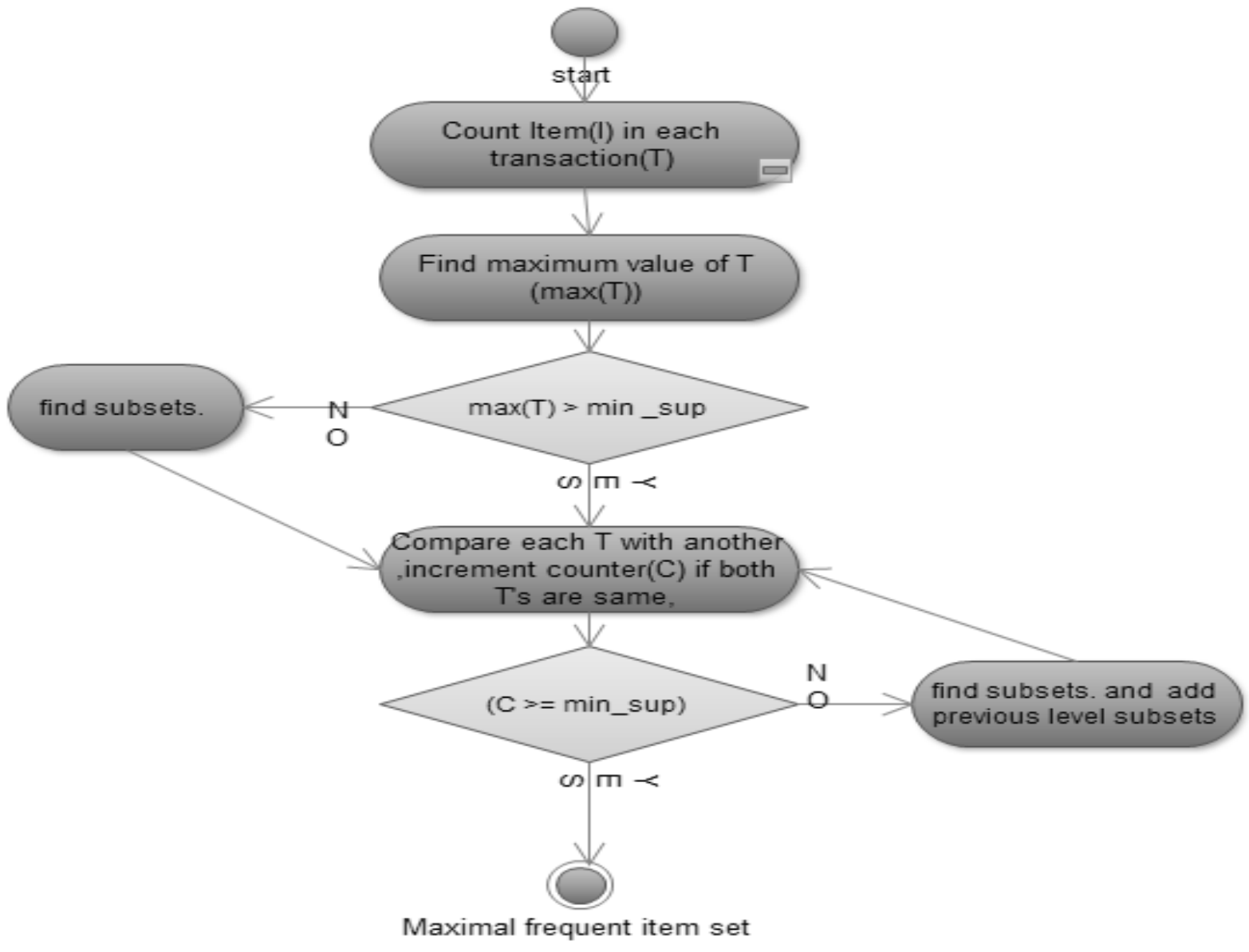

Figure 1: Procedure to find maximal frequent itemsets.

In this section we are presenting the proposed method to find maximal frequent itemsets. The working procedure is divided into 2 algorithms. The first is MFIF algorithm and the second is SUBSET FORMATION algorithm.

## MFIF ALGORITHM

```
Precondition:: i=0, max=0, Count=0;
   a [ ]← Count ( I ) in each T;
      // I represents Items //
      for (1 to n transaction)
          if( a[i]>max)
            max=a[i]
          endif
     endfor

  MOVE:
      if count(max) ≥ min_sup
              move max itemset to new_arr[ ][ ]
      endif
  FIND:
  for all transaction in new_arr[ ] [ ]
       Compare each I with (I-1) Items;
      If( I=(I-1) )
           Count++;
      endif
  endfor
  if(Count ≥ min_sup)
      Li ←All Itemsets with min_sup;
  else
      max=max-1;
  endif
   Create subsets of all transactions in
   new_arr[ ][ ] and store in sub_arr[ ][ ].
  goto MOVE;
     fin_arr[ ][ ]=new_arr[ ][ ]+sub_arr[ ][ ];
  goto  FIND;
```

## SUBSET FORMATION ALGORITHM

```
  for all transaction in new_arr[i][j]
            temp[k]=j;
              k++;
 endfor


        Initialize  l=l+i*4
  for all l less than or equal to k+i*4
              l++;
```

```
 endfor
 for all m items
        item[l][m]=new_arr[u][v];
      item[l][temp[w]]=0;
      w++;
      v=0;
endfor
Repeat until all the subsets are formed.
```

## 5. EXPERIMENTAL ANALYSIS

Fig.2 and Fig.3 shows the results of MFIF. The proposed method takes less time to find maximal frequent itemset. Fig.2 consists of 10 transactions of 20 items as input, in which two transactions have 12 items and the values are similar; it meets 20 percent of minimum support, hence 12 itemset results as most maximal frequent itemset. The experiment is done till 10000 transactions.

Another example, fig.3 consists of 10 transactions of 20 items as input, here only one transaction has 13 items, and count of 13 itemset transactions will be 1, which does not meet the minimum support. So subset formation is done. Subsets will be generated of 12 items from the transaction of 13 itemset, later the generated subsets and other transactions which have 12 items will move and combine in one array and compare the subsets .If count value is greater than minimum support, then that set will be the Maximal frequent itemset.

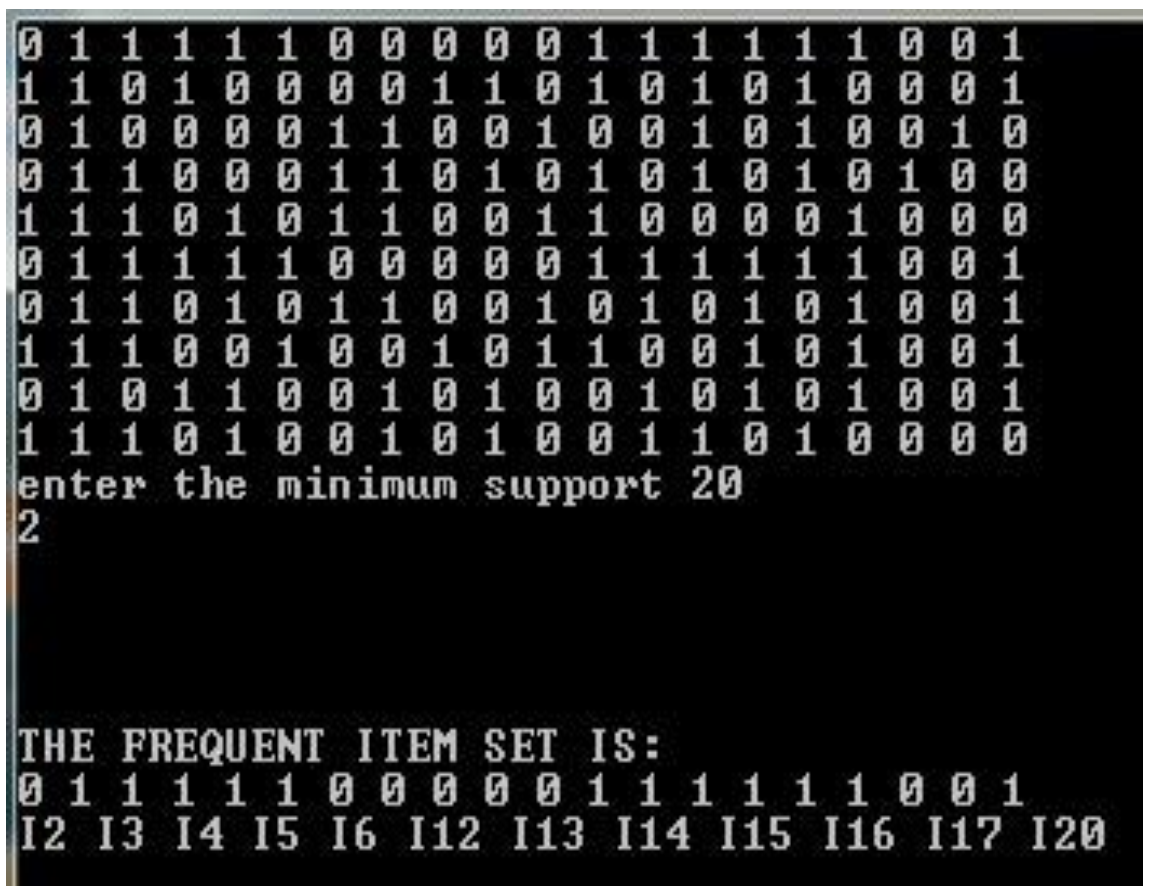

**Fig.2** MFIF Result: 12 itemset resulted as maximal frequent itemset.

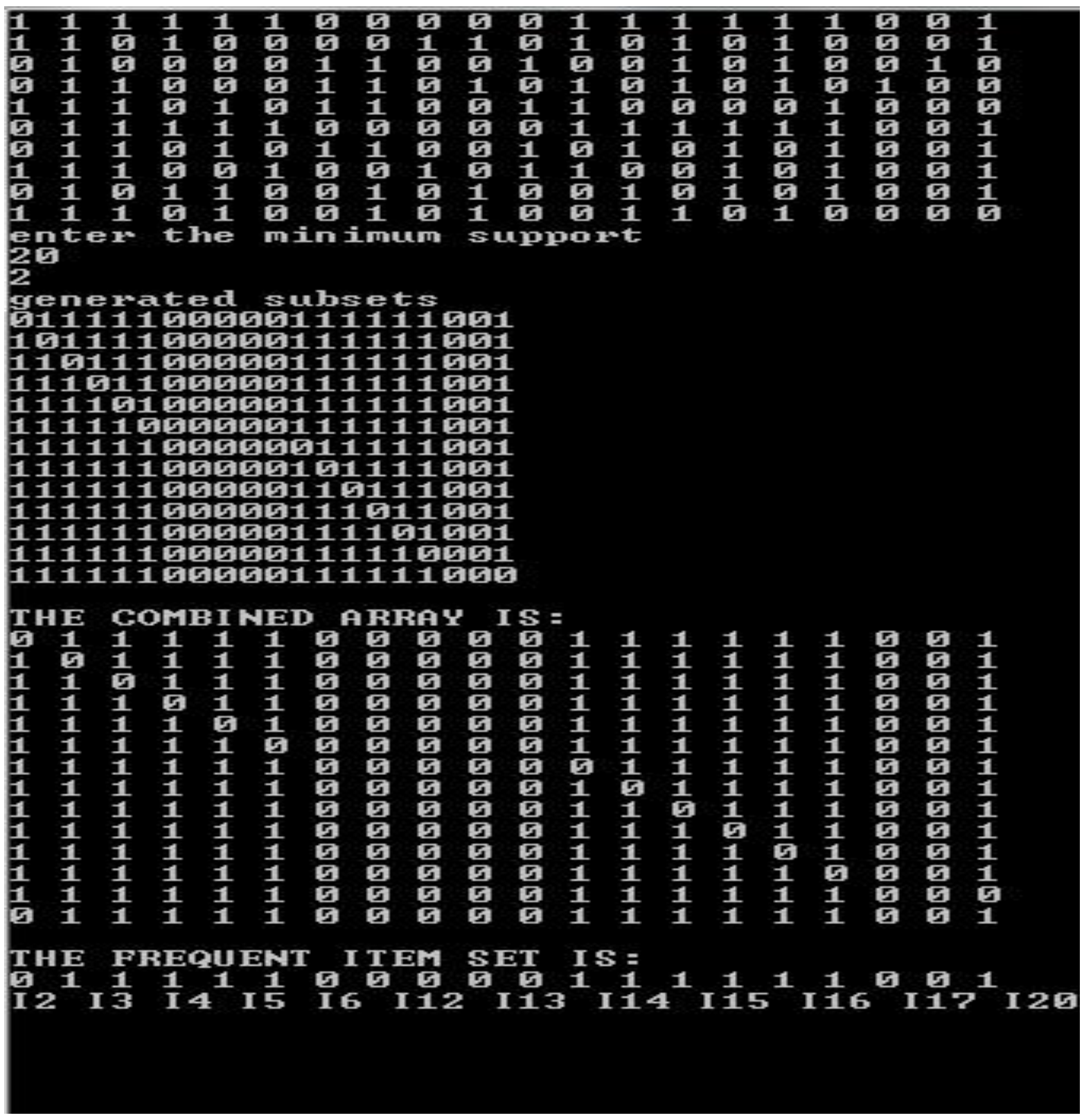

**Fig.3** MFIF Result: 12 itemset resulted as maximal frequent itemset with subset generation.

Complexity of the Apriori algorithm depends on the number of itemsets present in the transaction, i.e. if transaction has ‘n’ items, then we have to consider the items starting from 1 frequent itemset till we find out the ‘n’ frequent itemsets, so complexity increases as ‘n’ value increases.

MFIF(proposed algorithm) results in less time complexity compared to Apriori; when the itemsets are large, it does not depends on the value ‘n’. Complexity increases only at the generation of subsets of each itemsets, and yields less time complexity if maximal frequent itemset found at the initial stage.

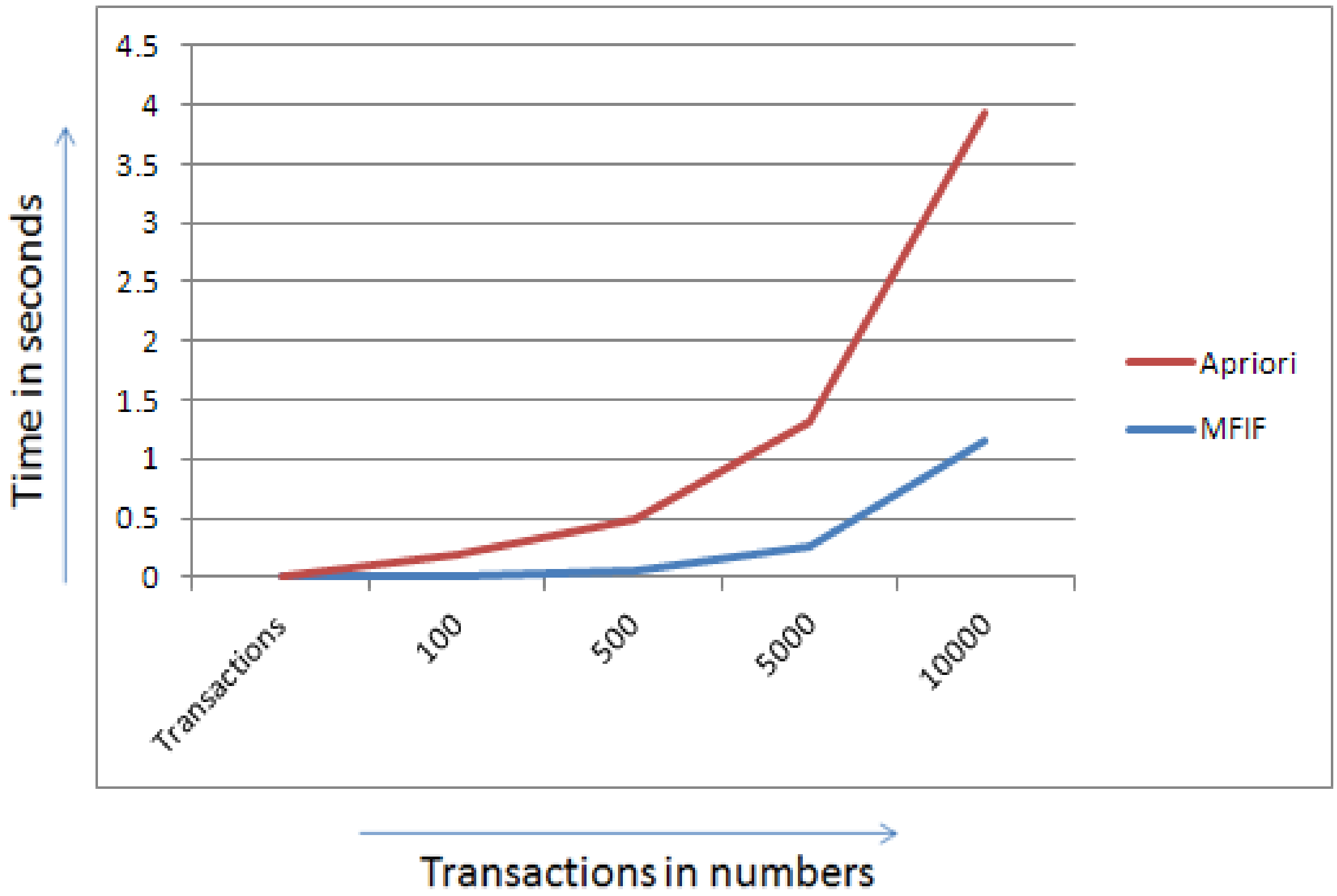

Fig.4. Graphical representation of time Complexity Comparison between Apriori and MFIF for frequent 12 itemsets.

| Transactions | MFIF (time in seconds) | Apriori (time in seconds) |
|---|---|---|
| 100 | 0.016 | 0.187 |
| 500 | 0.062 | 0.422 |
| 5000 | 0.266 | 1.047 |
| 10000 | 1.156 | 2.781 |

Table 1: Comparison between Apriori and MFIF

The results are shown in Table 1 and graphical comparison in fig.4. Time taken by MFIF and Apriori for 100, 500, 5000, 10000 transactions is shown in the table. Time complexity of MFIF is less than Apriori. Complexity of Apriori will increases as the number of items in the frequent itemset increases. In MFIF complexity does not depend on the number of itemsets present. But the time complexity increases only at the time of subset generation.

The results shows that, Apriori takes more time because it takes 12 scans to find out 12 element frequent set and MFIF takes 2 scans.

## 6. ADVANTAGES

- Too much memory space is not required for generation of subsets, because at a time only one level of element subsets are generated; as shown above only 12 element subsets are generated.
- Any element frequent set can be got in a single scan by subset creation method, which will help in applying any search method to traverse and get maximal frequent itemset, and it helps in reducing the scans drastically.

## 7. LIMITATIONS

- For scanning we assume that maximal frequent itemset will have at least 50 present of the total number of items present.
- Extra time is taken by the subset generation algorithm to calculate, but less time is taken in comparing the transactions as its just comparing whole transaction is equal.
- If items in Maximal frequent itemset are less, then MFIF algorithm takes more time than Apriori to calculate.

## 8. CONCLUSION

In data mining, association rule learning is a popular and well researched method for discovering interesting relations between objects in large databases. An efficient way to discover the maximal frequent set can be very important in some kinds of data mining problems .The maximal frequent set provides an effective representation of all the frequent itemsets. Discovering maximal frequent itemsets implies immediate discovery of all frequent itemsets. This paper presents a new algorithm that can efficiently discover the maximal frequent set. The top-down searching strategy is adopted in this algorithm. This approach can be very significant and effective to find maximal frequent itemset.